\documentclass[12pt]{article}
\usepackage{amssymb,amsfonts,amsmath,amsthm,psfrag}
\usepackage[dvips]{epsfig}

\newcommand{\AIC}{\text{AIC}}
\newcommand{\BIC}{\text{BIC}}

\begin{document}
\begin{center}
{\large\bf MIXED EFFECT MODELLING OF SINGLE TRIAL VARIABILITY IN ULTRA-HIGH FIELD fMRI}

{Christopher J. Brignell$^1$, William J. Browne$^2$, Ian L. Dryden$^1$ and Susan T. Francis$^3$}

{\it $^1$School of Mathematical Sciences, University of Nottingham, $^2$Graduate School of Education,
and Centre for Multilevel Modelling, University of Bristol and $^3$School of Physics and Astronomy, University of Nottingham}
\end{center}

\begin{quotation}
\noindent{\it Abstract:}
Neuronal brain activity in response to repeated stimuli can be perceived using functional magnetic resonance imaging (fMRI).  In this paper, we develop a statistical model for fMRI data that estimates both the associated haemodynamic response function and the within and between trial variability through maximum likelihood estimation.  We discuss our results in the context of other model-driven approaches, extending models already popular in the literature, while removing the need for some of their assumptions.  We consider an application to the motor cortex activity caused by a subject pressing a button and observe that the response changes significantly with task and through time.

\vskip 0.5cm

\noindent{\it Key words and phrases:}
Expectation-Maximisation algorithm, functional magnetic resonance imaging, Gaussian mixture model, haemodynamic response function, principal components analysis, random effects model, single trial variability, voxel classification.
\end{quotation}

{\bf 1. Introduction}

In response to a stimuli, the activation of a brain region can be studied using functional magnetic resonance imaging (fMRI).  Neural activity is detected by changes in the blood oxygenation level dependent (BOLD) haemodynamic response signal which is measured through monitoring magnetic resonance (MR) images of the subject's brain taken at regular intervals.  Psychologists and neuroscientists use fMRI experiments to identify MR signal intensity changes which correlate with the experimental paradigm presented to the subject.  Inference can then be made regarding the location and time-course of the underlying neural activity.

A typical fMRI experimental design consists of a number of trials, known as epochs.  During each epoch, the subject receives a stimulus or performs a task corresponding to one of the experimental conditions.  Each experimental condition is normally repeated several times and, for a large number of conditions, the experiment may be divided into a number of separate sessions.  Repeating the experiment on several subjects enables inferences to be made about the population.

At regular intervals during each epoch an MR scan gives a three-dimensional array of MR signal intensity measurements.  Each entry in the array corresponds to a three-dimensional pixel, known as a voxel, in the image.  The sequence of images taken during the experiment leads to a four-dimensional data set.  Physical constraints enforce a compromise between spatial and temporal resolution but modern scanners can typically record voxels of volume 10 mm$^3$ at intervals in the order of seconds.  A complete brain image consists of several hundred thousand voxels.

The aim of this paper is to build a statistical model for the haemodynamic response function (HRF) caused by a sequence of stimuli over a period of time and develop new inferential methods for fMRI data.  Brain activation is highly variable, and establishing how the brain's response changes to each stimulus (single-trial variability) is an important step to understanding the wider problem of how the different areas of the brain divide tasks and interact.

The primary motivation for this work is the analysis of images from an ultra-high field MR scanner.  The images are gradient echo planar images (EPI) taken at the Sir Peter Mansfield Magnetic Resonance Centre (SPMMRC), University of Nottingham, on the Philips 7T scanner.  Scanners with an ultra-high field of 7T have only become recently available.  The higher field increases the BOLD signal change and the signal to noise ratio, however, it also raises the confounding physiological noise levels.

The plan of the paper is as follows.  In Section 2 we introduce our case study and the experimental data acquired and examine previous work on this topic.  In Section 3 we develop our statistical model for fMRI data and maximise the likelihood using an Expectation-Maximisation (EM) algorithm.  In Section 4 we propose hypothesis tests for statistical inference on the parameter estimates.  We present results for the case study in Section 5.  Finally, we conclude with a brief discussion.

{\bf 2. Application and previous work}

{\bf 2.1 Application}

Many fMRI studies involve a physical response to a visual or audio stimulus, for example selecting an option from a keypad.  It is, therefore, vital for a wide range of applications that the time course of the BOLD signal change associated with neuronal activity, the haemodynamic response function (HRF), in the motor cortex can be modelled.  For this reason, we acquired images of a volunteer's motor cortex, which is an area of grey matter towards the posterior of the brain, that is responsible for controlling muscle movements.

A dynamic scan was taken at 2 second intervals and each dynamic scan was composed of 12 slices, each 3mm thick with a 0.7mm slice gap, taken sequentially at 1/6 second intervals.  Each slice is composed of 64 x 64 pixels of size 1mm x 1mm.  A total of ten visual stimuli were presented at 28.25 second intervals, which cued the volunteer to press a button once on odd-numbered trials or five times on even-numbered trials.

Acquired MR images are usually subjected to three preprocessing steps, namely slice-timing, realignment and smoothing, with the aim of removing confounding effects.  Slice-timing corrects for the time difference between each slice being recorded within each dynamic scan and transforms all the data within each scan to a single time point.  Realignment corrects for the small movement in the volunteer's position between each dynamic scan by translating and rotating each scan, in relation to the first, with a six-parameter rigid-body transformation.  Finally, the images are spatially smoothed to improve the signal-to-noise ratio and allow for any remaining imperfections in anatomical alignment, at the expense of spatial resolution.  The size of the spatial smoothing kernel should match the size of any potential activation we wish to detect.  Typically, this is 1.5 times the voxel size, and our images were smoothed with an isotropic Gaussian kernel of 1.5mm full-width half-maximum (FWHM).

\begin{figure}
\begin{center}
\includegraphics[angle=270,width=10cm]{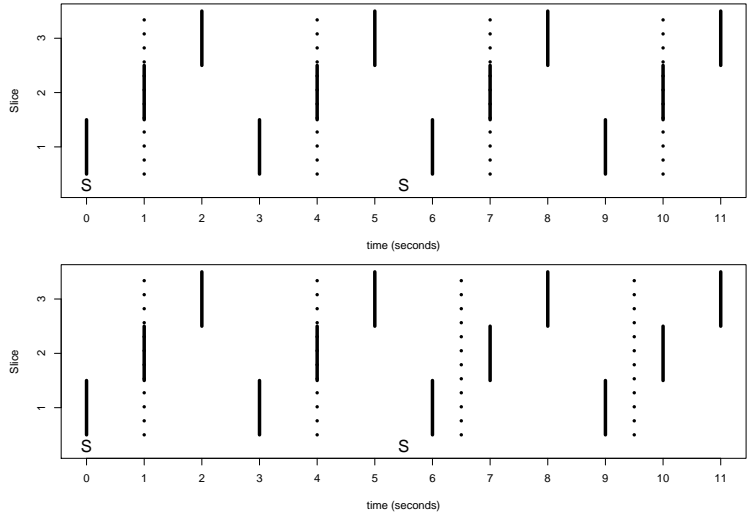}
\end{center}
\caption{Schematic diagrams showing the difference between slice timing (top) and slice/trial timing (bottom) for a simplified paradigm with a dynamic scan of 3 slices and a 5.5 second interval between stimuli.  Slice acquisition times are shown with a solid line, and the corrected timings with a dotted line.  The stimuli were given at times marked ``S''.}\label{fig:slicetime}
\end{figure}

We pre-processed our data using the SPM2 software package (Friston et al., 2002) but with one slight modification to the slice-timing.  The slice-timing algorithm models the time series from each slice as a linear combination of sinusoids of different phases and frequencies and the data is then shifted forwards or backwards in time by effectively adding a constant to the phase of every frequency.  Conventionally, each slice is shifted so that the time-series has the values that would have been obtained had the slice been acquired at the same moment as a reference slice.  However, due to the ``jitter'' asynchrony between the stimulus presentation times and the scan acquisition rate, we implement slice-timing with a different time shift for each slice/trial combination so that the slices are corrected to the same post-stimulus time points for each trial.  A schematic diagram of this can be seen in Figure \ref{fig:slicetime}.

During image preprocessing we automatically masked voxels at locations exterior to the brain such that they are not included in the following analysis.

{\bf 2.2 Background}

Current statistical methods for fMRI data analysis tend to focus on localising areas of the brain where the response changes between two experimental conditions.  The output is a classification in which each voxel is deemed active or inactive in processing the difference between the stimuli or tasks.  A careful choice of experimental conditions will enable a localisation of the neurons involved in processing precise brain functions.  There are two limitations to purely focusing on localising brain activity (Genovese, 2000).  Firstly, fMRI data is correlated in space and time due to the underlying biological process and the measurement technique.  For example, we would expect some correlation between voxels containing neurons performing similar tasks.  Similarly the haemodynamic response will be influenced by a voxel's proximity to major blood vessels, the location of which is unknown.  Secondly, localisation does not inform the experimenter about changes in the response between trials or the temporal dynamics of the process.

A variety of statistical techniques have been proposed for fMRI data.  These can be divided into data-driven and model-driven methods (Lu et al., 2005).  Data-driven methods include independent component analysis (e.g. Beckmann and Smith, 2004), principal components analysis (e.g. Hansen et al., 1999) and cluster analysis (e.g. Goutte et al., 1999).  The advantage of these techniques is the ability to identify differences in the HRF between conditions without specifying a hypothesis or a model.  Model-driven methods include the general linear model (e.g. Friston et al., 1995b) and the deconvolution model (Goutte et al., 2000).  Model-driven methods have gained in popularity due to the ease of interpretation and application.  For example, the general linear model (GLM) approach can be implemented using the software packages FSL (Beckmann et al., 2003), fMRIstat (Worsley et al., 2002) and SPM (Friston et al., 2002).

The majority of model-driven methods treat voxels independently, primarily for speed of computation.  In the GLM method, for example, a linear model is fitted with the voxel time series as the response variable and convolutions of a HRF with the experimental paradigm as covariates.  The corresponding parameters, one for each condition, and contrasts of parameters can then be estimated at each voxel and plotted over a brain image to form a statistical parametric map (SPM).  Hypothesis tests, such as t-tests and ANOVAs, with a null hypothesis of no activity can then be performed to localise voxels with significant activity.

\begin{figure}
\begin{center}
\includegraphics[angle=0,width=10cm]{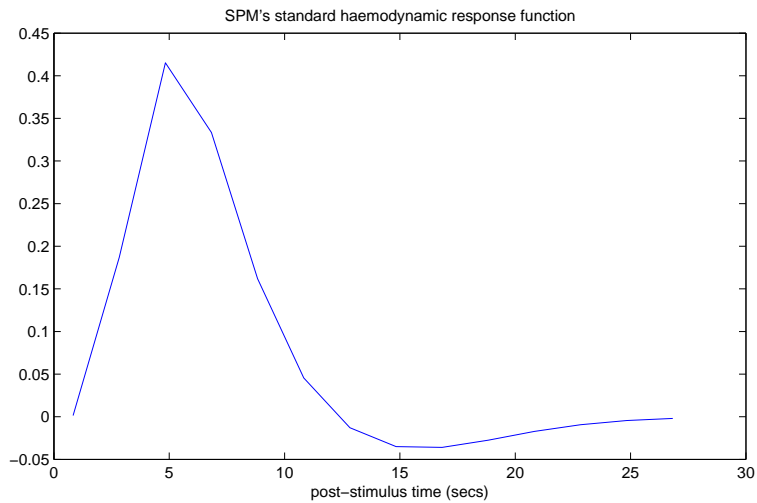}
\end{center}
\caption{SPM2's standard haemodynamic response function evaluated at two second intervals.}\label{fig:spmhrf}
\end{figure}

A basic implementation of the GLM method assumes the HRF shape is known.  The standard canonical HRF employed by SPM2 (Worsley and Friston, 1995), for example, is shown in Figure \ref{fig:spmhrf}.  Its shape is derived from prior empirical evidence, demonstrating a positive peak 4-8 seconds after the onset of activity and a negative undershoot at the return to baseline, and is formed by the discretised summation of two curves taken from a gamma probability density function.  The disadvantage of assuming a fixed shape is that the statistical tests have less power if the shape is incorrect.

One solution to this problem is to increase the number of basis functions in the linear model.  Studies have tried expanding the number of basis functions based on derivatives with respect to time and dispersion (Friston et al., 1998), principal component analysis of a large number of HRFs (Woolrich et al., 2004b), terms of a Fourier series (Josephs et al., 1997) and cosine functions (Zarahn, 2002).  Taking the number of basis functions to the extreme, we could estimate the HRF shape by including one free parameter for every sampled time point, as implemented in the finite impulse response (FIR) model (Glover, 1999).  The disadvantage of these latter methods is that there are lower degrees of freedom and less power if the covariates are not orthogonal.

A number of studies have tried more flexible models to estimate the HRF shape from the data.  For example, the HRF shape has been parameterised as the discretised summation of inverse logit functions (Lindquist and Wager, 2007) or through cubic splines (Genovese, 2000).  The disadvantage of these methods is that the parameters can usually only be estimated through non-linear algorithms or simulation techniques, such as Markov chain Monte Carlo (MCMC), which may have convergence problems and be computationally expensive.  Other approaches include the selective averaging of responses to each experimental condition (Dale and Buckner, 1997).  From these estimates of HRF shape, inferences can then be made regarding the onset, peak latency and response duration.  Both the GLM models and the more flexible models assume that some of the HRF parameters differ between voxels, although sometimes only the magnitude parameter, but not between trials.

The assumption of independent trials is usually made because the BOLD signal change is of the order of 2.5\% and the signal-to-noise ratio (SNR) is low.  Averaging over trials, assuming that the response to each stimulus is dependent only on location and time since the last stimulus, improves the SNR.  In truth, this is probably a false assumption.  Experiments have shown that the HRF varies between voxels (Aguirre et al., 1998) and between epochs (Duann et al., 2002).  Behavioural studies have shown ``warm up" (Eysenck and Frith, 1977) and ``carry over" (Ward et al., 2001) effects, where peak performance is not achieved when starting a new task from rest or when switching immediately from one task to another.  If a subject makes a mistake in one trial, there will be a systematic effect on the behaviour in the next trial (Debener et al., 2005).  Duann et al. (2002) revealed ``dramatic and unforeseen haemodynamic response variations not apparent to researchers analysing their data with event-related response averaging and fixed haemodynamic response templates".

Adapting fMRI data analysis techniques to model trial-to-trial variability within a voxel is non-trivial.  In the GLM context, it can be done using a non-additive analysis of variance (ANOVA), where the time since the stimulus and the epoch number are taken as the treatment group and block effect respectively (Auffermann et al., 2001).  An epoch-specific parameter scales the treatment effect, corresponding to a change in amplitude over trials, and the parameters are estimated using maximum likelihood estimation (MLE).  Alternatively, an extension of the FIR model estimates a separate HRF function for each experimental condition, which is allowed to vary in amplitude over trials within the condition (Hinrichs et al., 2000).  The HRFs and the amplitudes are estimated through a non-linear optimisation algorithm assuming a Toeplitz-type covariance structure that is constant over voxels.

Devising a model which allows more complicated variation between trials greatly increases the number of parameters.  For example, an alternative extension of the FIR model is to estimate an HRF for each trial but constrain the estimate to lie in a neighbourhood of the original HRF estimate using non-linear optimisation techniques (Lu et al., 2005).  Another approach is to model the response to each trial using splines with parameters for lags and amplitudes (Genovese, 2000).  This model is very flexible and, if prior distributions are specified for the parameters, the posterior distribution can be maximised or sampled from using MCMC.

Our approach will be to use a linear mixed effects model with an overall HRF, with one free parameter for each time point, a single fixed effect scale parameters for each voxel and, finally, random effects for errors within and between epochs.  Our approach provides a flexible model with a manageable number of parameters.

{\bf 3. Modelling the data}

\begin{figure}
\begin{center}
\includegraphics[angle=270,width=10cm]{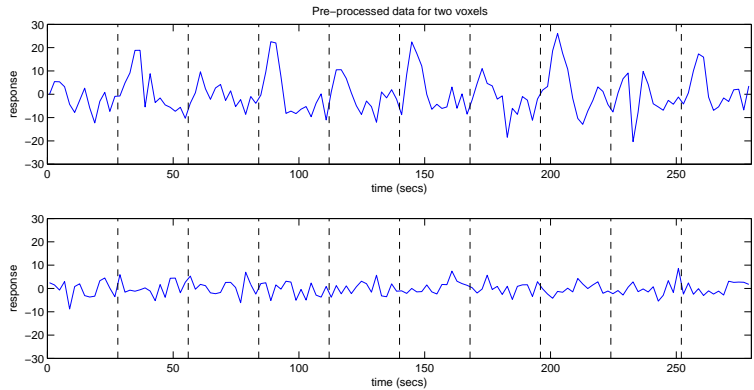}
\end{center}
\caption{The pre-processed data from two example voxels.  The stimuli were presented at the times marked with dotted lines.  One shows signs of activation (top), whilst the other is mainly noise (bottom).}\label{fig:data}
\end{figure}

Let $\widetilde{Y}(t)$ be the observed MR signal from a specific voxel at time $t=1,\dots,N$, where $N$ is the total number of images in the experiment.  The MR signal intensity is affected by covariates such as movement, blood flow and oxygen levels.  Therefore, let $\widetilde{X}$ be the corresponding $N \times q$ design matrix, composed of image preprocessing and physiological data, with the mean of each column being zero.  The response variable and the design matrix are both subjected to temporal filtering using the first few non-constant basis functions of a one-dimensional cosine transform to remove low-frequency confounds, such as MR scanner drift.  This is a standard SPM2 preprocessing step.  The mean MR signal intensity is also subtracted from the data to set the voxel's baseline signal to zero.  Following this transformation, we denote the design matrix as $X$ and let $Y_i$ be the haemodynamic response of the $i$th voxel.  The pre-processed data for two voxels are shown in Figure \ref{fig:data}.

Let $E$ be the number of epochs (or trials) in the experiment and let $T$ be the number of images during each epoch (i.e. $N = T \times E$).  Let $h(t)$ be an unknown vector of length $T$ that represents the haemodynamic response in each epoch.  We expect the response at each stimuli to be similar, so we let $\mu = 1_E \otimes h$ represent a convolution between $h(t)$ and stimulus onset.  Let $\Sigma_E$ and $\Sigma_T$ denote variability between and within events, respectively, where the subscript denotes the dimension of the covariance matrix.  We cannot estimate $\mu$, $\Sigma_E$ and $\Sigma_T$ for individual voxels due to the number of parameters involved, so we introduce some regularisation across voxels.  Figure \ref{fig:data} shows that not all voxels within the brain are active during an experiment, so we base the estimates on active voxels only.  This is similar to SPM2's method for estimating variability based on voxels demonstrating significant regression (Mumford and Nichols, 2006).  We assume, however, no prior information regarding which voxels are active.  Therefore, we formulate a multivariate Gaussian mixture model for two classes of activity (active and inactive).

We model the response for voxel $i$, $Y_i$, with
\begin{eqnarray*}
\text{active component:} && Y_i \sim N_{TE}( \mu_1 = \beta_i \mu + X b_i, \Sigma_1 = \Sigma_E \otimes \Sigma_T ) \text{ and}\\
\text{inactive component:} && Y_i \sim N_{TE}( \mu_2 = X b_i, \Sigma_2 = \sigma^2 I_n ).
\end{eqnarray*}
where $\beta_i$ denotes the scale of the response at the $i$th voxel, $b_i$ is the parameter vector for the design matrix and $\sigma^2$ is a scalar.  To remove the arbitrary scaling of the response parameter we constrain the haemodynamic response function to have unit norm, i.e. $\| h \| = 1$.  Note that the model for the active component is the mixed effects model, $Y_{ijk} = \beta_i h_k + (X b_i)_{jk} + u_{ijk}$, $j=1,\dots,E$, $k=1,\dots,T$, where
\begin{eqnarray*}
\left( u_{i11},\dots,u_{iET} \right)^T \sim N \left( 0, \Sigma_E \otimes \Sigma_T \right),
\end{eqnarray*}
$\beta_i$, $h$ and $b_i$ are fixed effects and $u_{ijk}$ are the random effects.  The Gaussian mixture model is then $f(Y_i) = pf_1(Y_i) + (1-p)f_2(Y_i),$ where $f_1 \sim N(\mu_1, \Sigma_1)$ and $f_2 \sim N(\mu_2, \Sigma_2)$.  The log-likelihood for the data is,
\begin{eqnarray*}
l(p,\mu_1,\mu_2,\Sigma_1,\Sigma_2 | Y) = \sum_{i=1}^V \log \left[ p f_1(Y_i | \mu_1, \Sigma_1) + (1-p) f_2(Y_i | \mu_2, \Sigma_2 \right)],
\end{eqnarray*}
where $V$ is the number of voxels.

Maximising this likelihood function is non-trivial.  A solution does exist if we are prepared to constrain $\Sigma_1 = \Sigma_2$ (Day, 1969) but this requires the false assumption that both classes of activity demonstrate the same covariance structure.  An alternative solution is to assume each voxel is either active or inactive and then evaluate the likelihood for all possible voxel allocations.  This is equivalent to minimising $|\hat{\Sigma}_1|^{V_1} |\hat{\Sigma}_2|^{V_2}$, where $\hat{\Sigma}_i$ is the MLE of the covariance matrix for the $V_i$ voxels in group $i=1,2$.  However, there are $2^{V-1}$ possible combinations, which is computationally impractical.

Instead, we utilise the Expectation-Maximisation (EM) algorithm (Dempster et al., 1977) to maximise the likelihood and estimate the model parameters from the data.  The algorithm requires that the data is augmented with latent variables, which are effectively missing data.  Let $Z_i$ be a Bernoulli random variable with $P(Z_i=1)=p$.  Let $Y_i \sim f_1$ if $Z_i=1$ and $Y_i \sim f_2$ if $Z_i=0$.  The joint density of $(Z_i,Y_i)$ is
\begin{eqnarray*}
f(Z_i,Y_i) = \left[ pf_1(Y_i) \right]^{Z_i} \left[ (1-p)f_2(Y_i) \right]^{1-Z_i},
\end{eqnarray*}
and the log likelihood for the data is,
\begin{eqnarray*}
\lefteqn{ l(p,\mu_1,\mu_2,\Sigma_1,\Sigma_2 | Z,Y) = }\\
&& \sum_{i=1}^V Z_i \log \left[ pf_1(Y_i | \mu_1, \Sigma_1) \right] + (1-Z_i) \log \left[ (1-p)f_2(Y_i | \mu_2, \Sigma_2) \right].
\end{eqnarray*}

The EM algorithm is an iterative procedure with each iteration composed of expectation and maximisation steps.  The expectation step estimates the values of the latent variables and then the conditional log-likelihood is maximised by differentiation.  Let $\Theta' = (p,\mu_1,\mu_2,\Sigma_1,\Sigma_2)$ be the current parameter estimates and let $\Theta$ denote the parameter values at the next iteration then,
\begin{eqnarray*}
\lefteqn{ Q(\Theta | \Theta', Y) = \mathbb{E} \left[ l(\Theta|Z,Y) |Y,\Theta' \right], }\\
&& = \sum_{i=1}^V \mathbb{E}(Z_i|Y,\Theta') \log \left[ \frac{pf_1(Y_i|\Theta)}{(1-p)f_2(Y_i|\Theta)} \right] + \log \left[ (1-p)f_2(Y_i|\Theta) \right],
\end{eqnarray*}
where $Q$ has its usual EM interpretation and,
\begin{eqnarray*}
\mathbb{E}(Z_i|Y,\Theta') = f(Z_i=1|Y,\Theta') = \frac{pf_1(Y_i|\Theta)}{pf_1(Y_i|\Theta)+(1-p)f_2(Y_i|\Theta)} = p_i.
\end{eqnarray*}
Due to the small values of $f_1$ and $f_2$, it is computationally more feasible to calculate $p_i = 1/(1+e^c)$, where $c= \log(1-p) - \log(p) + \log(f_2) - \log(f_1)$.

We write the function, $Q$, in terms of our model parameters,
\begin{eqnarray*}
Q(\Theta | \Theta', Y) &=& \sum_{i=1}^V p_i \log \left[ pf_1(Y_i | \mu_1, \Sigma_1) \right] + (1-p_i) \log \left[ (1-p)f_2(Y_i | \mu_2, \Sigma_2) \right], \\
&=& -\frac{VET}{2} \log\pi + \sum_{i=1}^V \Big\{ p_i \log(p) - \frac{p_i E}{2} \log |\Sigma_T| - \frac{ p_i T}{2} \log |\Sigma_E|\\
&& - \frac{p_i}{2} (Y_i - \mu \beta_i - X b_i)^T \Sigma_1^{-1} (Y_i - \mu \beta_i - X b_i)\\
&& + (1-p_i) \log (1-p) - \frac{(1-p_i) E T}{2} \log (\sigma^2)\\
&& - \frac{(1-p_i)}{2\sigma^2} (Y_i - X b_i)^T (Y_i - X b_i)\Big\}.
\end{eqnarray*}

The model parameters can be estimated in turn by iteratively maximising $Q$ conditional on the current values of the other parameters.  The conditional MLEs are given below, with their derivation provided in the appendix,
\begin{eqnarray*}
\hat{p} &=& \frac{1}{V} \sum_{i=1}^V p_i, \\
\hat{\beta}_i &=& (\mu^T \Sigma_1^{-1} \mu)^{-1} \mu^T \Sigma_1^{-1} (Y_i - X b_i), \\
\hat{b}_i &=& (A^T A)^{-1} A^T \left( p_i X^T \Sigma_1^{-1} (Y_i - \mu \beta_i) + (1 - p_i) X^T \Sigma_2^{-1} Y_i \right), \\
\hat{h} &=& \frac{ \left(\sum_i \sum_j \sum_k p_i e_{jk} \beta_i (Y_{ij} - X_j b_i) \right) }{ \left( \sum_i p_i \beta_i^2 \right) \left( \sum_j \sum_k e_{jk} \right) }, \\
\hat{\Sigma}_T &=& \frac{1}{E \sum_{i=1}^V p_i} \sum_{i=1}^V p_i R_i \Sigma_E^{-1} R_i^T, \\
\hat{\Sigma}_E &=& \frac{1}{T \sum_{i=1}^V p_i} \sum_{i=1}^V p_i R_i^T \Sigma_T^{-1} R_i, \\
\hat{\sigma}^2 &=& \frac{1}{ET \sum_{i=1}^V (1-p_i)} \sum_{i=1}^V (1-p_i) (Y_i - X b_i)^T (Y_i - X b_i).
\end{eqnarray*}
where $A = p_i X^T \Sigma_1^{-1} X + (1 - p_i) X^T \Sigma_2^{-1} X$, $Y_{ij}$ and $X_j$ are the rows of $Y_i$ and $X$, respectively, corresponding to event $j$, $e_{jk}$ is the $j,k$th entry of $\Sigma_E^{-1}$ and $R_i$ is the $T \times E$ matrix of residuals in the active model.

EM algorithms are sensitive to the starting estimates of the model parameters.  For our application we initially set $h$ to be the standard haemodynamic response function shown in Figure \ref{fig:spmhrf} and set $\Sigma_E$ and $\Sigma_T$ to be identity matrices.  A reduced version of the model that assumes all voxels are active was then fitted.  The model parameters for the reduced model can be estimated using a simpler version of the algorithm presented above but with no expectation step.  t-tests of the resulting $\beta_i$ parameters provided an initial classification of active and inactive voxels for the EM algorithm.  This classification was used to produce starting estimates of the proportion of active voxels, $p$, and the covariance matrices for the two groups, $\Sigma_1$ and $\Sigma_2$.

{\bf 4. Statistical Inference}

Our choice of model suggests that active voxels will have $p_i \approx 1$ and a large positive $\beta_i$ parameter.  Active voxels, therefore, can be identified by testing the null hypothesis, $H_0: \beta_i = 0$, versus the alternative hypothesis, $H_1: \beta_i > 0$, for $i = 1, ..., V$ under the model, $Y_i \sim N_{TE} (\mu \beta_i + X b_i, \Sigma_1)$.  A transformation gives $Y_i^\ast \sim N_{TE} (\mu^\ast \beta_i + X^\ast b_i, I_{TE})$ where $Y_i^\ast = \Sigma_1^{-1/2} Y_i$, $\mu^\ast = \Sigma_1^{-1/2} \mu$ and $X^\ast = \Sigma_1^{-1/2} X$.  The test statistic under $TE-q-1$ degrees of freedom is,
\begin{eqnarray}
T_i = \frac{\hat{\beta}_i}{(S_i^2 ({\mu^\ast}^T \mu^\ast)^{-1})^{1/2}} \sim t_{TE-q-1}, \label{eq:ttest}
\end{eqnarray}
where $S_i^2 = \| Y_i^\ast - \mu^\ast \beta_i - X^\ast b_i \|^2 / (TE-q-1)$.

There has been much debate in the neuroimaging community regarding how to set the statistical significance level in fMRI studies (Marchini and Ripley, 2000), due to problems in estimating the magnitude of the response and the correlation in spatially and temporally correlated data.  SPM2 utilises set-level inference, using distributional approximations from the theory of Gaussian random fields (Friston et al., 1996).  The method assesses the probability of obtaining $c$ or more clusters, containing $v$ or more voxels.  Our voxel-by-voxel approach is the simplest case of this, allowing clusters of just one voxel.   The use of Gaussian random fields has potential advantages over SPM2's previous methods (Friston et al., 1994; Friston et al., 1995a; Worsley and Friston, 1995), which corrected the significance level for temporal correlations only.  Our method of pre-whitening the data, using the estimated covariance matrix to transform from $Y$, $\mu$ and $X$ to $Y^\ast$, $\mu^\ast$ and $X^\ast$, is similar to that of Worsley et al. (2002), although they restrict the covariance matrix to be of auto-regressive form.  We correct for multiple comparisons by adjusting the initial p-value threshold of 0.001 to control the false discovery rate (Benjamini and Hochberg, 1995) using the adaptive method (Benjamini and Hochberg, 2000).

The goal of single-trial variability analysis is to identify trends in the way active voxels respond in space and time to stimuli.  Here a key ingredient of our work is to examine the random effects at the active voxels.  We carry out this task by investigating the principal components scores of the fitted random effects covariance matrices at the active voxels.  Principal components (PC) analysis of $\Sigma_T$ will highlight key differences in the response of voxels to different events.  The $k$th PC score for voxel $i$ at event $j$ is,
$$s_{ijk} = \gamma_k^T (Y_{ij} - \beta_i \mu - X_j b_i),$$
where $\gamma_k$ is the $k$th PC loading.  We model the $k$th PC score using an analysis of variance with event number as a categorical variable.

{\bf 5. Results}

In this section we apply statistical models to the motor-cortex data in Section 2.  The post-stimulus times chosen for our study were $\frac{5}{6},2\frac{5}{6},4\frac{5}{6},\dots,26\frac{5}{6}$ seconds, as these correspond to the middle-slice acquisition times during the first trial, so $T=14$. There were $E = 10$ epochs, so $N = T \times E = 140$.  We choose $q=6$, using the six realignment parameter vectors as covariates because physiological data were unavailable.  Recall that $X$ is the $N \times q$ design matrix.  The model proposed in Section 3 has a much more complicated structure than the simpler linear model currently implemented by SPM2, so we compare 5 models of varying complexity.  The first two linear models we consider are,
\begin{eqnarray*}
\textrm{Model 1:} && Y_i \sim N_{TE}( \beta_i \mu_s + X b_i, \sigma^2 I_{TE} );\\
\textrm{Model 2:} && Y_i \sim N_{TE}( \beta_i \mu + X b_i, \sigma^2 I_{TE} );
\end{eqnarray*}
where $\mu_s$ is the HRF used by SPM2 and $\mu$ is an HRF estimated from the data.  In both models the shape of the HRF is constant for each voxel and trial but the magnitude of the response varies at each voxel through the estimation of $\beta_i$, $i=1,...,V$.  We compare these to Gaussian mixture models where the active component is modelled by,
\begin{eqnarray*}
\text{Model 3: Active component} && Y_i \sim N_{TE}( \beta_i \mu + X b_i, \Sigma_E \otimes I_T ) ;\\
\text{Model 4: Active component} && Y_i \sim N_{TE}( \beta_i \mu + X b_i, I_E \otimes \Sigma_T ) ;\\
\text{Model 5: Active component} && Y_i \sim N_{TE}( \beta_i \mu + X b_i, \Sigma_E \otimes \Sigma_T ) ;
\end{eqnarray*}
and, in each case, the inactive component is modelled by $Y_i \sim N_{TE}( X b_i, \sigma^2 I_{TE} )$.  In these models the expected response and covariance matrix structure is different for the active and inactive components, and the models only differ in the complexity of the covariance structure for the active component.

For models 3-5, the EM algorithm was run until the change in the norm of parameter estimates between successive iterations fell below a specified tolerance of 0.0001.  The EM algorithm took approximately 50 iterations to converge for the model and data used in this study.  The algorithm only takes a few minutes to compute but it will require more iterations if the amount of data or covariates is increased.

\begin{figure}
\begin{center}
\includegraphics[angle=270,width=6cm]{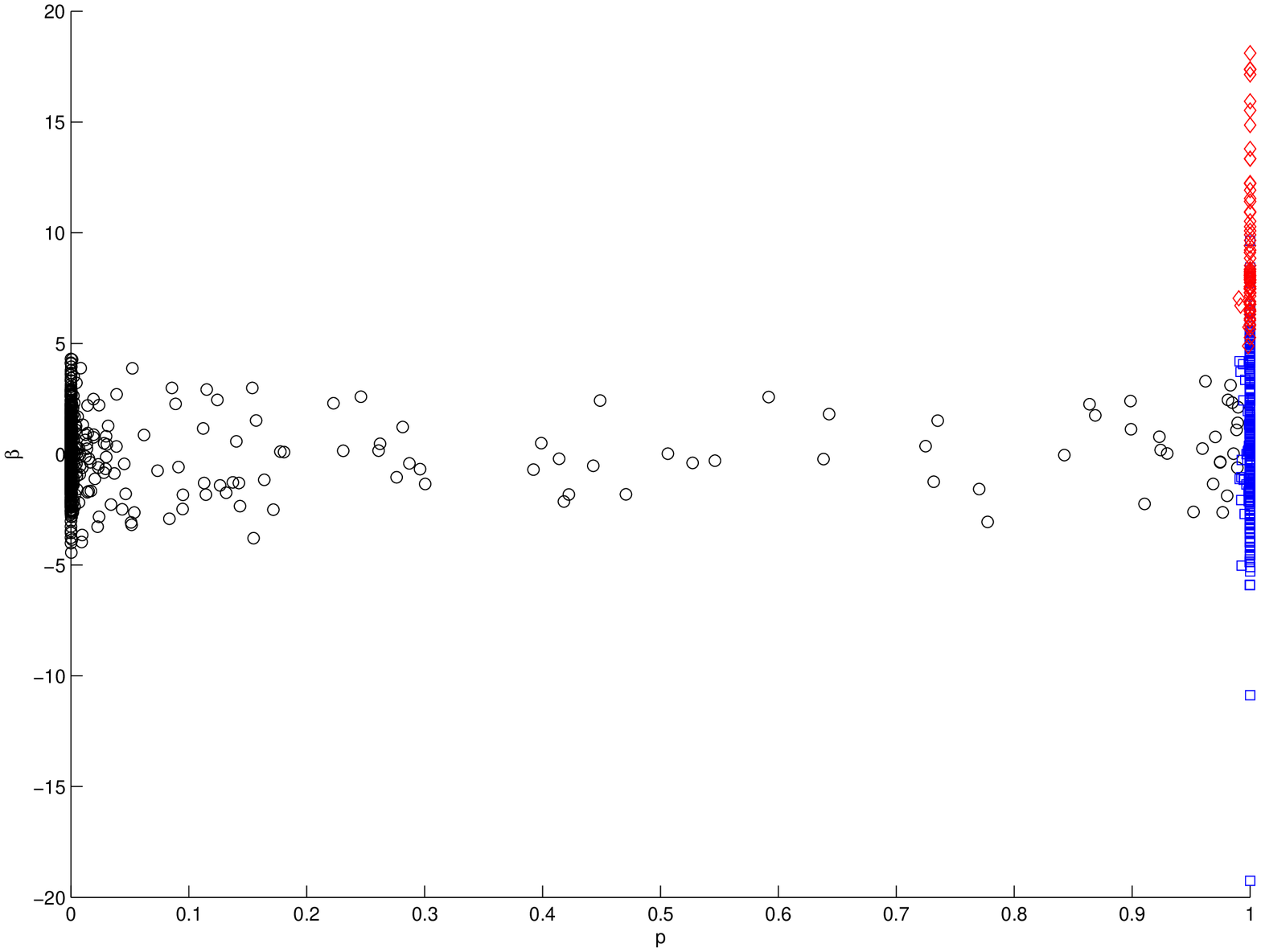}
\includegraphics[angle=270,width=6cm]{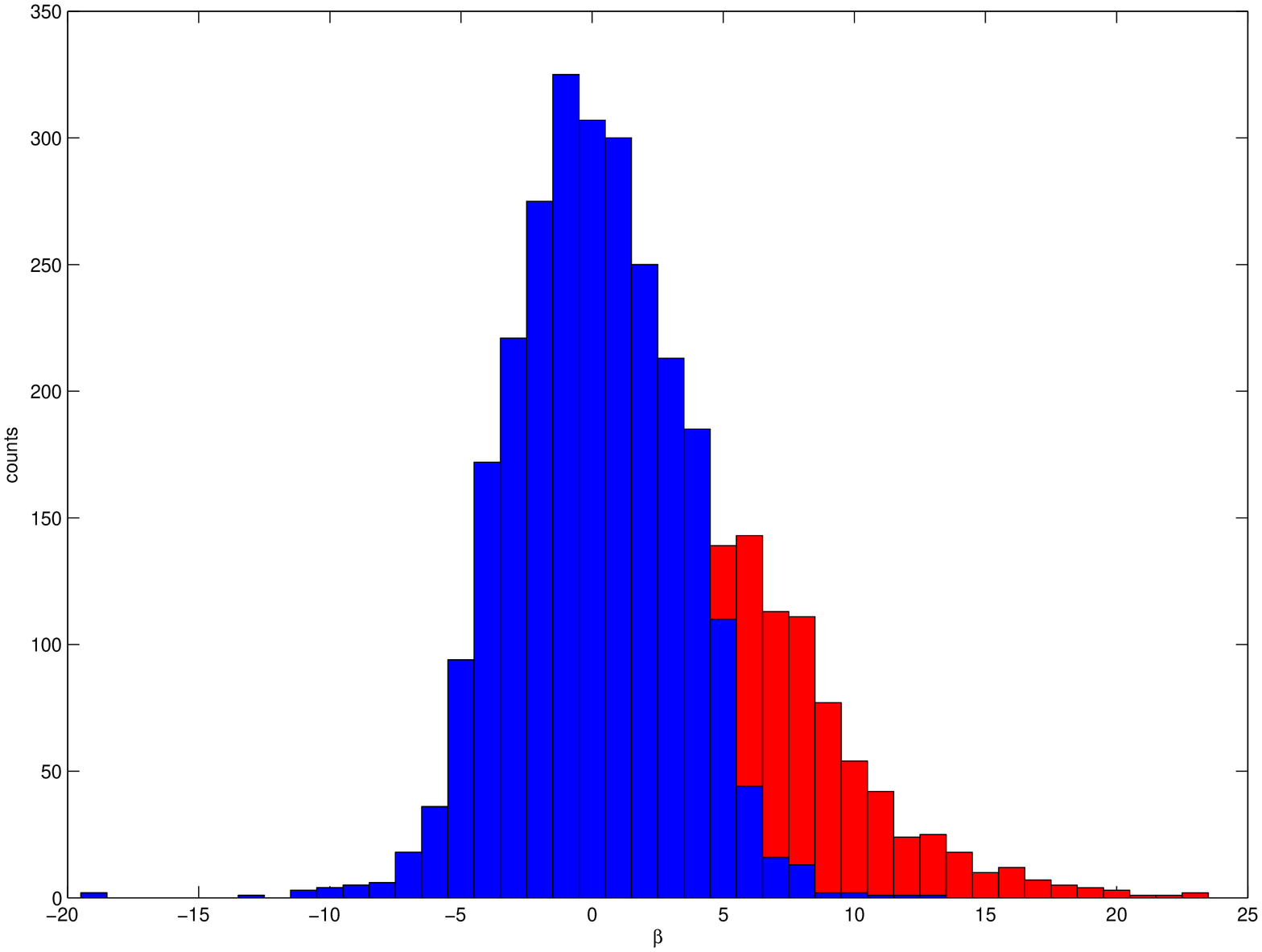}
\end{center}
\caption{A scatter plot of $p_i$ versus $\beta_i$ (left) and a histogram of $\beta_i$ values (right), showing active voxels (red) and voxels that share their covariance structure (blue).}\label{fig:betaptilde}
\end{figure}

The Akaike information criterion (AIC) and Bayesian information criterion (BIC) were calculated for each of the five models using the log-likelihood, $\log L$, the number of model parameters, $P$, and the number of observations, $n$.  The results are shown in Table \ref{table:modelcomparison}.  The more complicated models are favoured by a low AIC and BIC due to the addition of relatively few parameters.  The reduction from model 1, based on SPM2 analysis, to model 2 highlights the benefit of allowing the data to estimate the haemodynamic response.  Similarly, the large increase in the log-likelihood from model 2 to models 3-5 shows the need for separating the active and inactive components with a mixture model.  A mixture model allows the variability of active responses to be estimated using only an appropriate subset of the voxels.  The reduction in AIC and BIC between model 5 and model 4 suggests an improvement in the relative model fit and gives evidence that responses to sequential epochs are correlated and should not be treated as independent responses.  Based on Table \ref{table:modelcomparison}, it seems reasonable to model the data using model 5.

\begin{table}[ht]
\begin{center}
\caption{The deviance and AIC for five models.}\label{table:modelcomparison}
\vskip 10pt
\begin{tabular}{|c|c|c|c|c|}
\hline
Model & $P$ & $ \log L$ & $\AIC$ & $\BIC$ \\\hline
1 & 70434 & -12481601 & 25104069 & 25940417 \\
2 & 70447 & -12348551 & 24837996 & 25694502 \\
3 & 80566 &  -3898027 &  7957185 &  8936720 \\
4 & 80616 &  -3894722 &  7950676 &  8931818 \\
5 & 80671 &  -3889950 &  7941242 &  8922053 \\
\hline
\end{tabular}
\end{center}
\end{table}

A scatter plot of $p_i$ versus $\beta_i$ for model 5 is shown in Figure \ref{fig:betaptilde}.  It highlights that voxels generally fell into three categories.  Firstly, voxels with $p_i \approx 1$ and $\beta_i$ significantly greater than zero by the t-test in Equation (\ref{eq:ttest}), are the voxels which are truly active.  Secondly, voxels with $p_i \approx 1$ and with $\beta_i$ not significantly greater than zero are voxels which share a covariance structure with the active voxels but do not demonstrate a large haemodynamic response.  Finally, voxels with $p_i \approx 0$ have an MR signal that consists mostly of white noise.  This provides some justification for using the Gaussian mixture model since truly active voxels nearly always have $p_i \approx 1$ and we have removed some of the influence of voxels containing noise from the estimates of variability between and within trials.

\begin{figure}
\begin{center}
\includegraphics[angle=0,width=12cm]{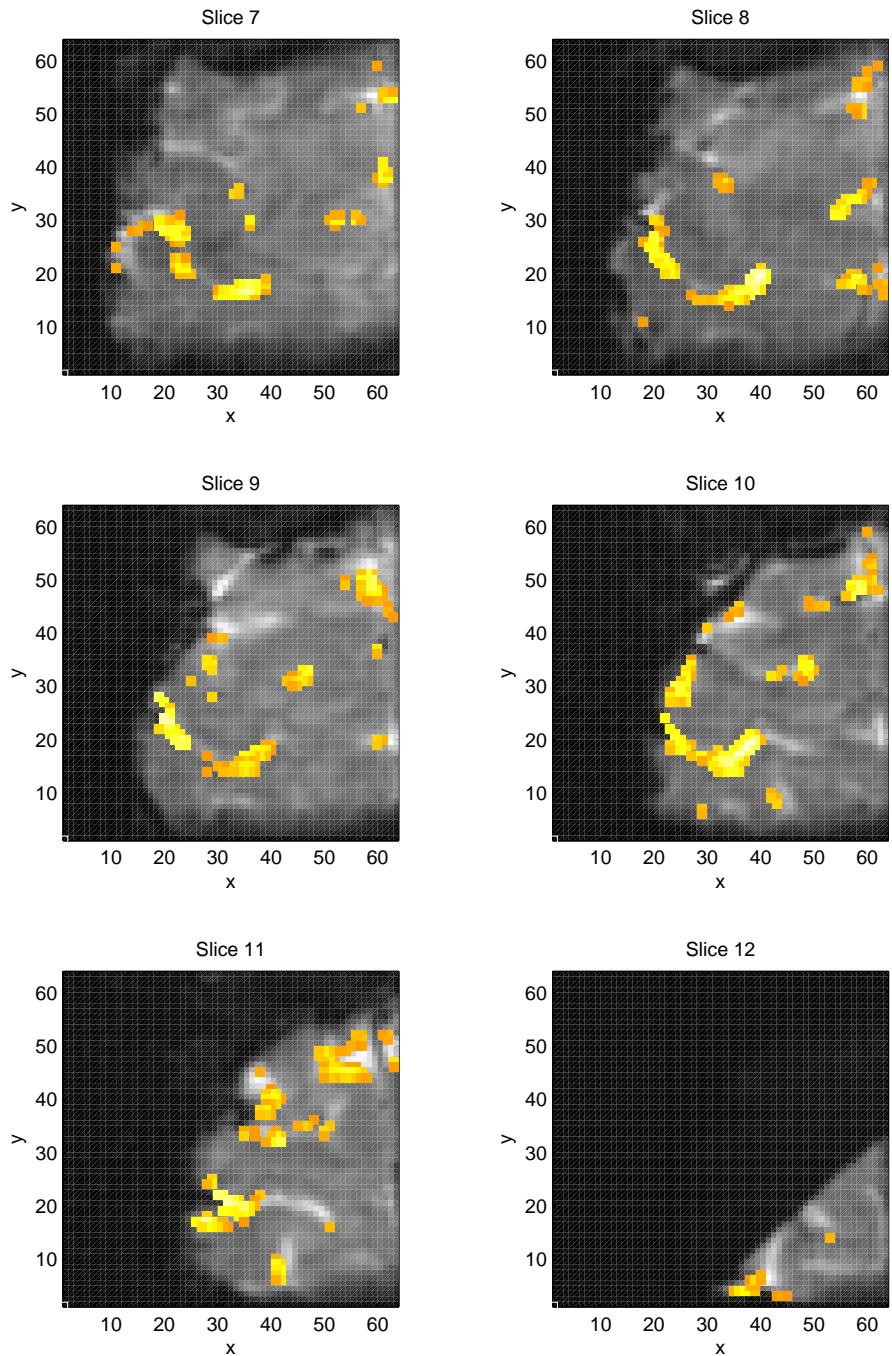}
\end{center}
\caption{The locations of active voxels in the motor cortex.  Lighter colours indicate increased significance.}\label{fig:map}
\end{figure}

\begin{figure}
\begin{center}
\includegraphics[angle=0,width=6cm]{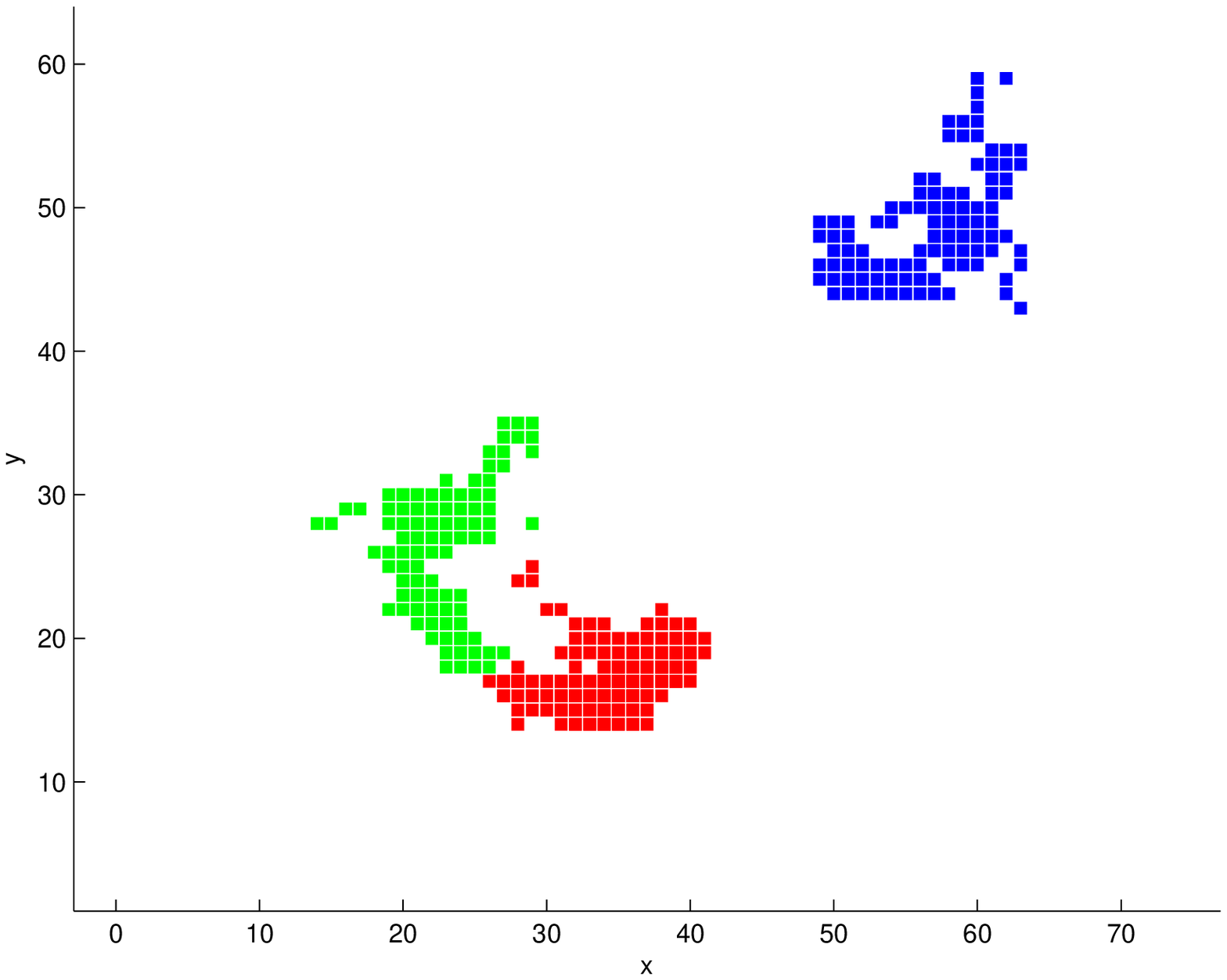}
\end{center}
\caption{The location of active clusters, projected into the $x$-$y$ plane.}\label{fig:clusters}
\end{figure}

The voxels where $\beta_i$ is significantly greater than zero are highlighted on the activation map in Figure \ref{fig:map}.  Note that head movement during scanning has caused part of the image in slice 12 to not be consistently recorded throughout the experiment and consequently this portion of the image is masked during analysis.  Slices 1-6 are not displayed for clarity and because they contain less activation.  Cluster analysis of the voxel co-ordinates revealed three large groups of voxels, colour coded in Figure \ref{fig:clusters}.  These areas of activation are similar to those detected using SPM2 with the simplest model, see Section 6, but here they illustrate where single trial variability is assessed.

\begin{figure}
\begin{center}
\includegraphics[angle=0,width=12cm]{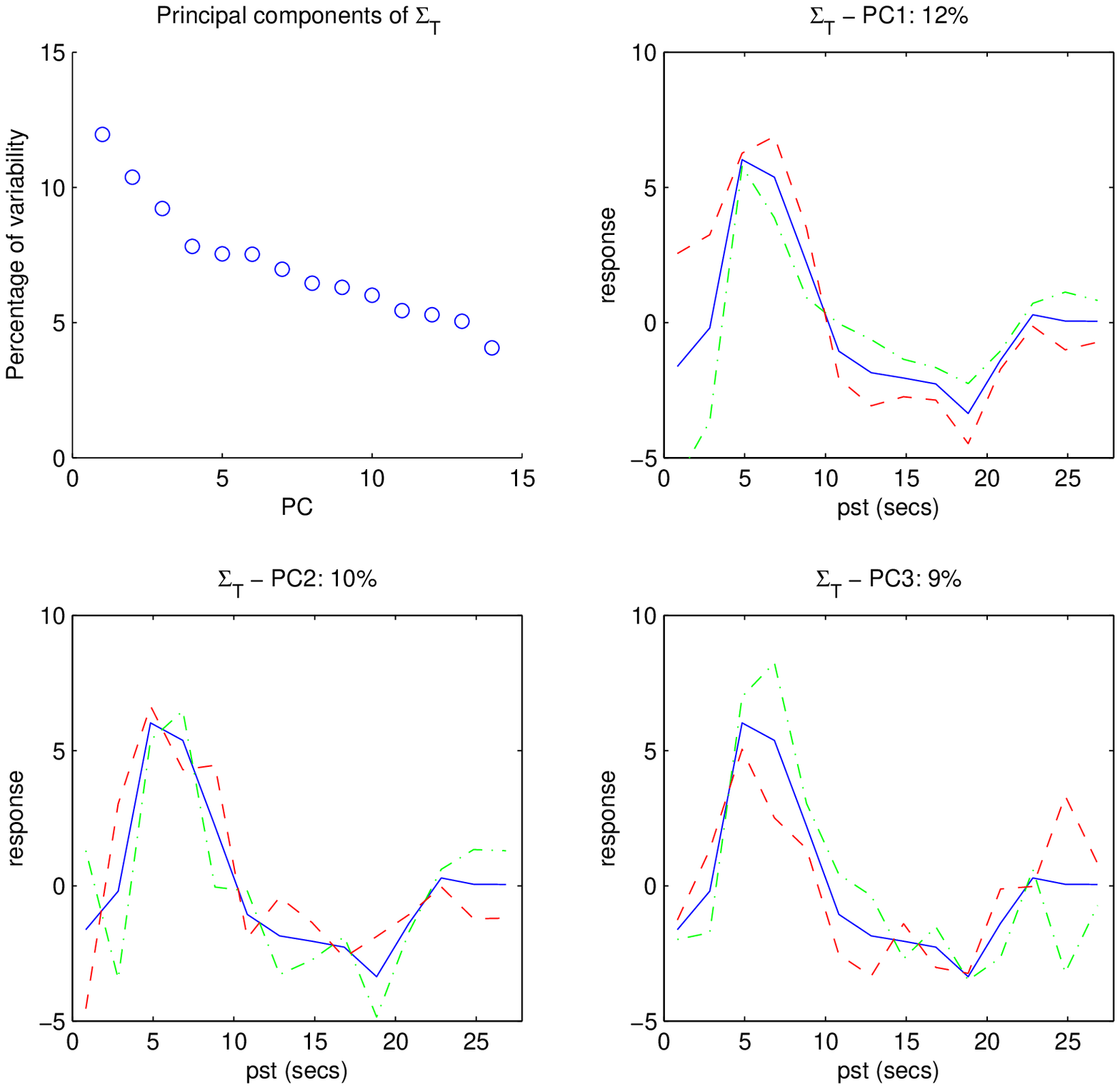}
\end{center}
\caption{The first plot shows the percentage of variability in each principal component of $\Sigma_T$.  The remainder show 10 times the mean response plus(green)/minus(red) the first three principal components of $\Sigma_T$.}\label{fig:SigmaT}
\end{figure}

\begin{figure}
\begin{center}
\includegraphics[angle=270,width=9cm]{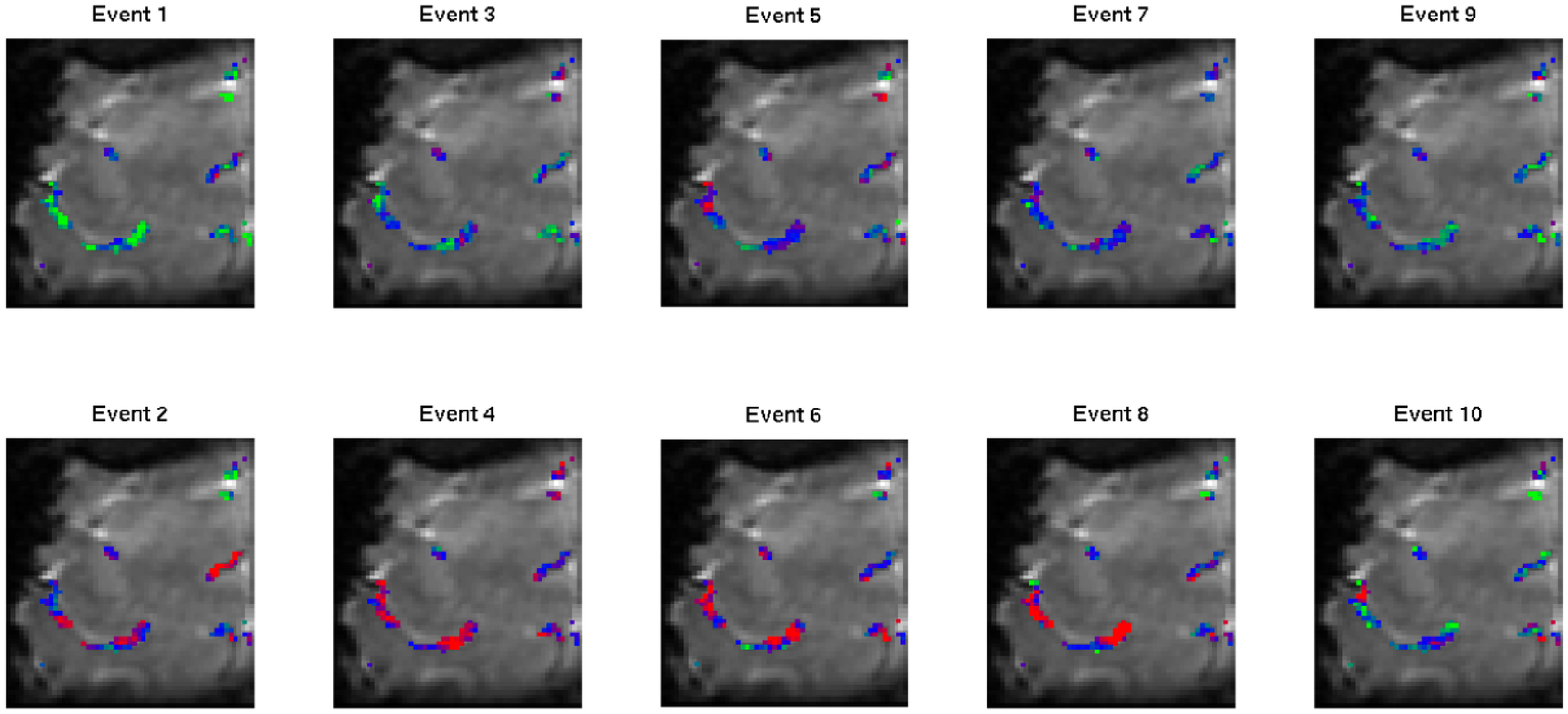}
\end{center}
\caption{PC1 scores for voxels in Slice 8 for the ten different events.  Odd-numbered trials required single button-presses (top row) and even-numbered trials required multiple button-presses (bottom row).  PC scores ranged from -10 (red), through zero (blue) to +10 (green).}\label{fig:pcmap}
\end{figure}

Figure \ref{fig:SigmaT} shows that the first three principal components of $\Sigma_T$ explain a litle more variability than the others.  The plots show the mean response, $\beta \mu$, and the effect of the first three principal components of $\Sigma_T$ on the mean, $\beta \mu \pm \lambda_k^{1/2} \gamma_k$, where $\beta=10$ and  $\lambda_k$, $\gamma_k$ are the $k$th eigenvalue and eigenvector of $\Sigma_T$.  Voxels with a low PC1 score show a much earlier rise in the response and a slightly later decay.  The spatial distribution of PC1 scores for Slice 8 is shown in Figure \ref{fig:pcmap}.  PC2 accounts for the width of the response, with low scores corresponding to a longer period of activation.  Voxels with a high PC3 score have a much later peak and a stronger response.

\begin{figure}
\begin{center}
\includegraphics[angle=0,width=12cm]{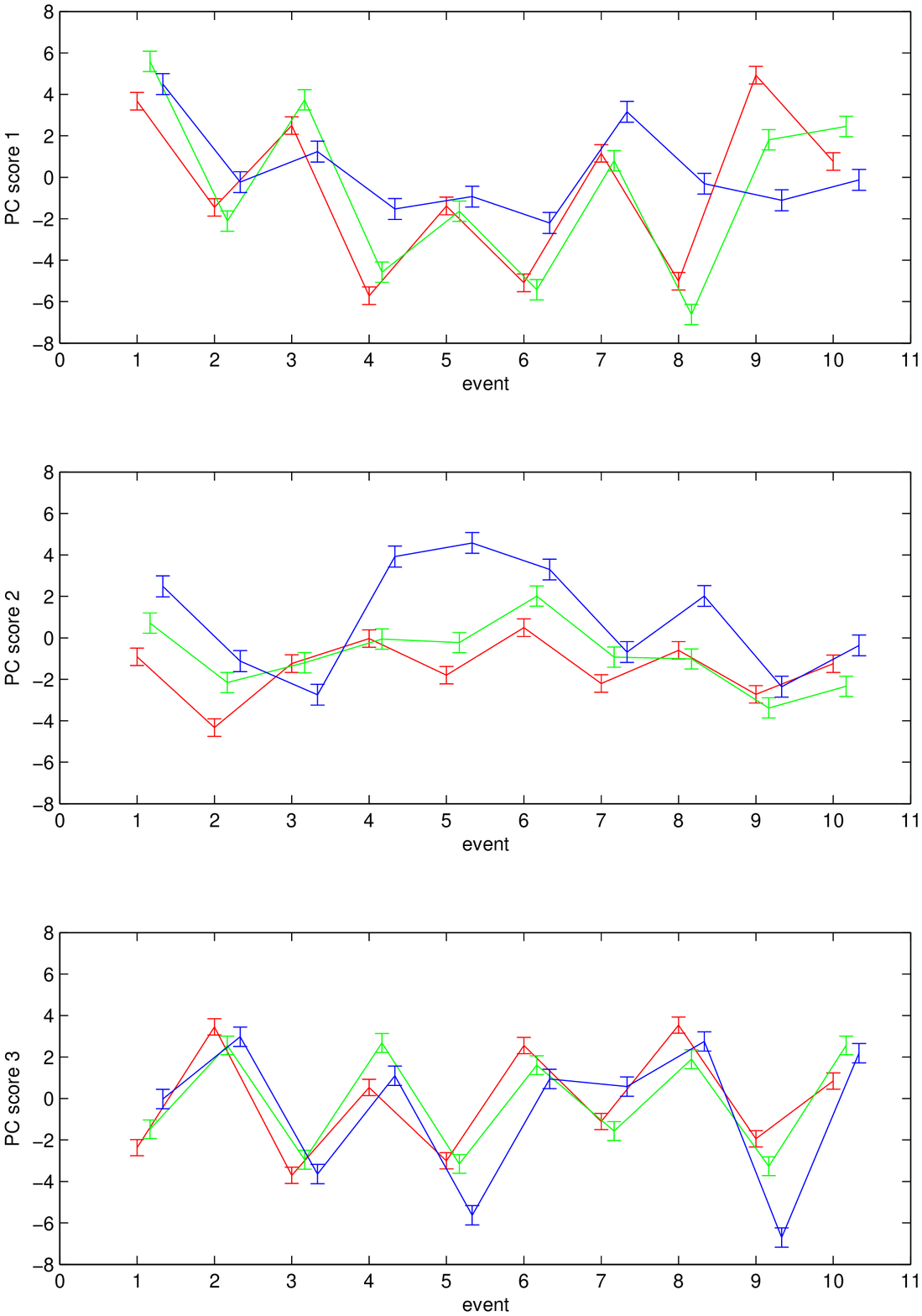}
\end{center}
\caption{The fitted values and standard errors from the ANOVA analysis of PC scores.  The colours correspond to cluster 1 (red), cluster 2 (green) and cluster 3 (blue), as shown in Figure \ref{fig:clusters}.  The results for each cluster have been offset to give greater clarity.}\label{fig:pcmodel}
\end{figure}

The PC scores were modelled using a two-way analysis of variance (ANOVA) with event number and cluster as categorical factors.  The fitted values from the ANOVA model are shown in Figure \ref{fig:pcmodel}.  Both PC1 and PC3 display a strong association with event type.  There are more similarities between cluster 1 and 2 responses compared to the cluster 3 response.  This mirrors the spatial relationship of these clusters and could reflect differences in each region's function in processing the stimuli or task.  For example, both motor and somatosensory areas are involved in this task and are captured in the MR image.

\begin{figure}
\begin{center}
\includegraphics[angle=0,width=12cm]{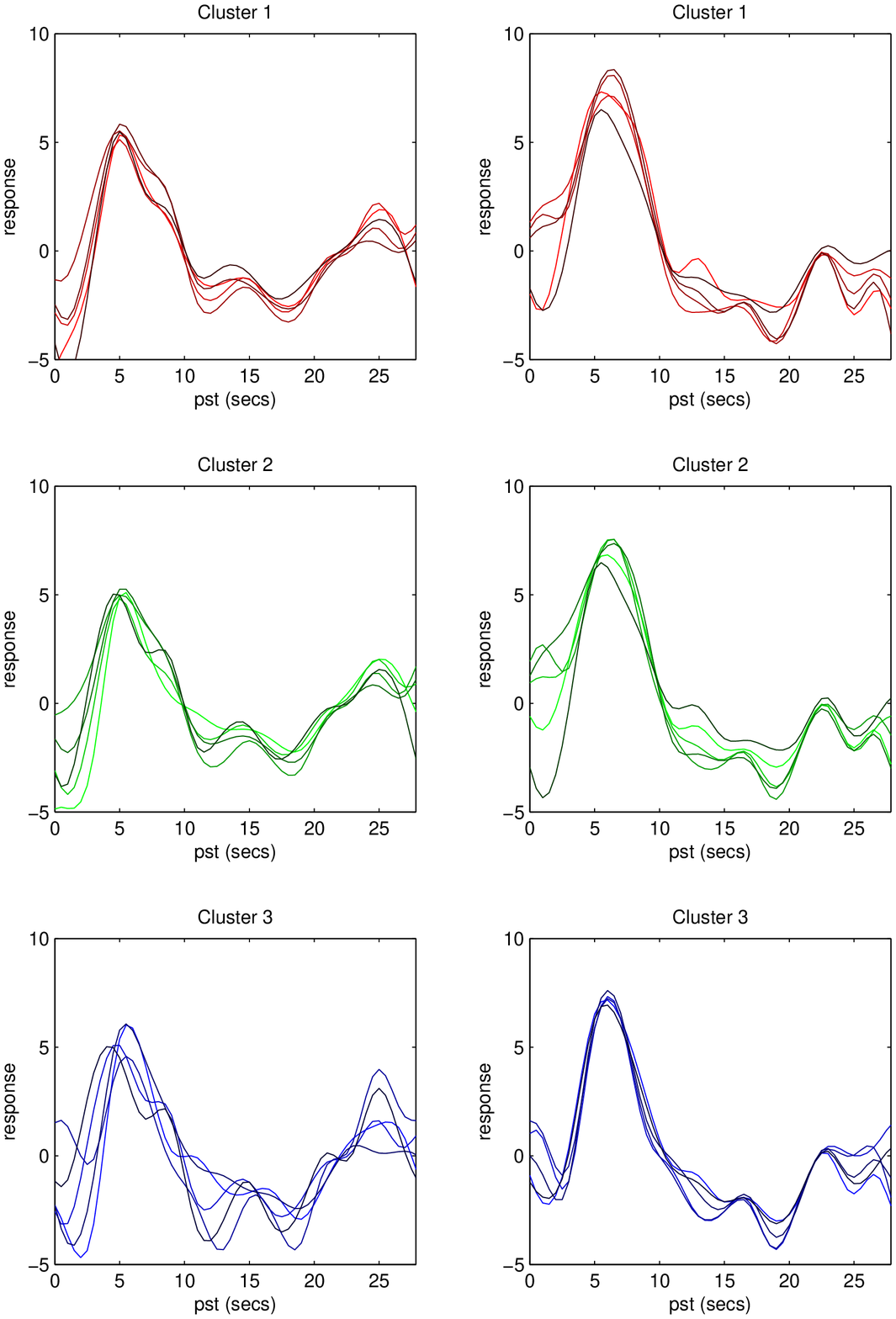}
\end{center}
\caption{The fitted haemodynamic response for the three clusters for one-press trials (left column) and five-press trials (right column).  Individual trials are represented by different shades, with darker colours indicating later trials.}\label{fig:fittedvalues}
\end{figure}

Changes in haemodynamic response for a particular cluster can also be seen through plotting the fitted values,
\begin{eqnarray*}
\hat{Y}_{cj} = \beta_c \mu + \sum_{k=1}^3 \hat{s}_{cjk} \gamma_k,
\end{eqnarray*}
where $\beta_c$ is the mean value of $\beta_i$ for the voxels in cluster $c$ and $\hat{s}_{cjk}$ are the fitted PC scores from the ANOVA analysis.  Figure \ref{fig:fittedvalues} shows the fitted values which have been interpolated using a cubic smoothing spline to aid interpretation.  The plots clearly show that for clusters 1 and 2 the response is much stronger and peaks later with a larger undershoot in trials where the volunteer presses the button multiple times.  This effect is less pronounced in cluster 3.  However, it should also be noted that the one-press trials generally start from a lower baseline caused by a prolonged undershoot in the preceding multiple-press trials.  There is a tendency for the response to be largest in the middle of the experiment.

A principal component analysis of $\Sigma_E$ revealed that the loadings do not display much structure.  However maximum likelihood estimation of $\Sigma_E$ shows a relatively large increase in the likelihood compared to $\Sigma_E = I_E$ and so it is worthwhile including this term in the model.

{\bf 6. Discussion}

\begin{figure}
\begin{center}
\includegraphics[angle=0,width=12cm]{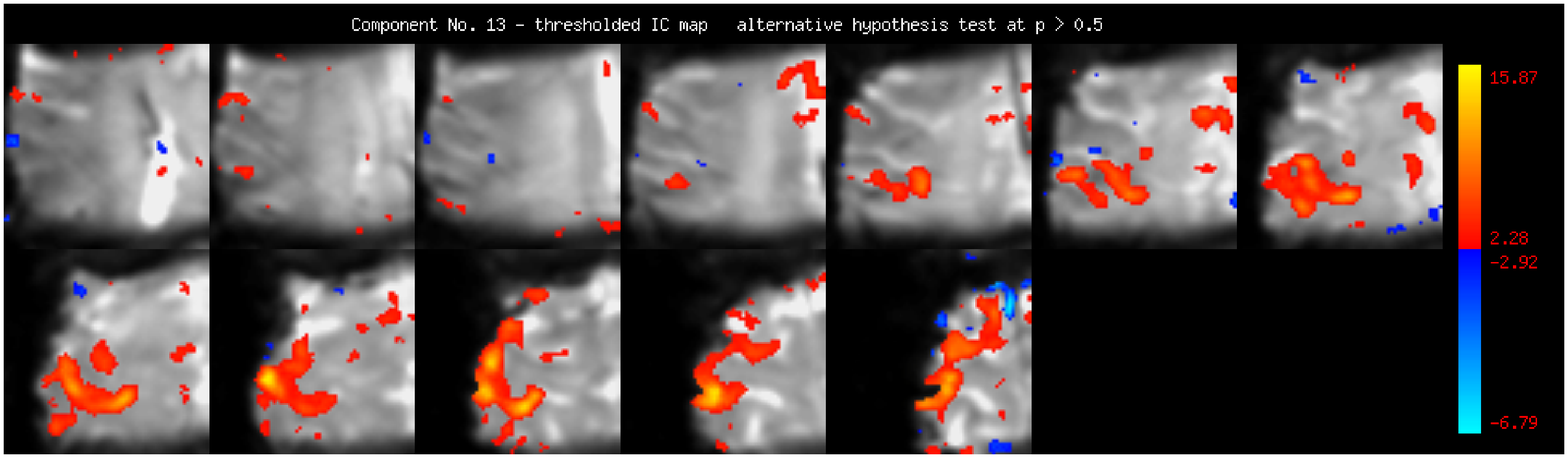}
\end{center}
\caption{Highlighted voxels indicate locations where the signal is present.}\label{fig:ICA}
\end{figure}

\begin{figure}
\begin{center}
\includegraphics[angle=90,height=6cm]{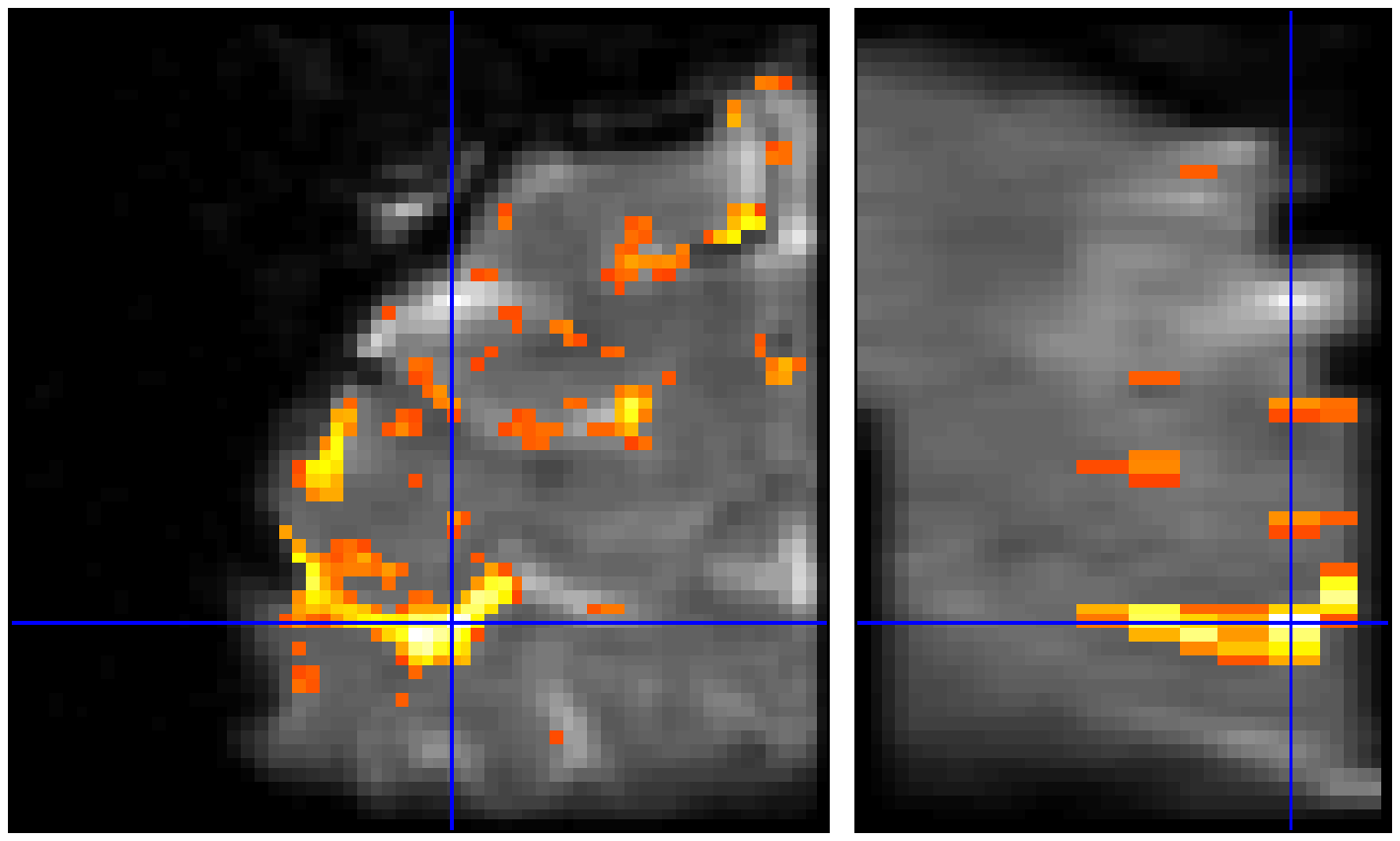}
\includegraphics[angle=0,height=6cm]{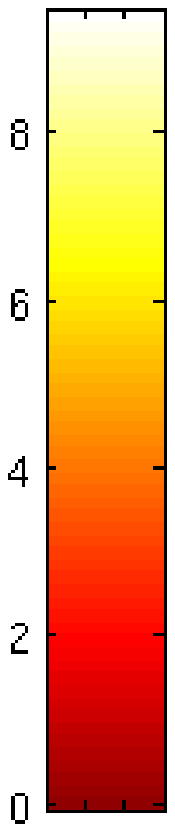}
\end{center}
\caption{Highlighted voxels have significant non-zero activity under SPM2's model with a p-value below 0.001.}\label{fig:spm}
\end{figure}

In this paper we have developed a statistical model for analysing single trial variability in fMRI data.  The chosen model produced a significantly higher likelihood than other simpler models, see Table \ref{table:modelcomparison}, including those implemented by software packages currently available.  For comparison, we analyse the same data using two current statistical software packages, MELODIC (Beckmann and Smith, 2004) which implements a probabilistic independent components analysis, and SPM2 (Friston et al., 1995b), which fits a linear model at each voxel, with a design matrix composed of covariate information and a basis function of a typical haemodynamic response.  These methods are data and model driven, respectively, and their resulting activation maps are shown in Figures \ref{fig:ICA} and \ref{fig:spm}.  Unsurprisingly, the active areas for both these methods have similar location to the active voxels detected using our model.  We can conclude, therefore, that our model indicates areas of the brain active in performing the task of pressing a button, which concur with other methods of analysis.

The key advantage of our method, however, is that we can also obtain maximum likelihood estimates of the shape of the response and the within and between trial variability.  The three main sources of variation, displayed through principal components analysis, were found to be the timing of the initial rise in response, the length of the response and the strength of the peak.  The analysis of principal components scores through a second-level model is similar to the multi-level models used in analysing between subject variability.  fMRI group analysis for a particular voxel combines a low-level fixed effects model, similar to model 1 above, for the within-subject analysis with a high-level random effects model for the between-group analysis (Woolrich et al., 2004a).

For example, the low-level model for the $i$th voxel in the $k$th subject is $Y_{ik} = X_k \beta_{ik} + \epsilon_{ik}$, where $Y_{ik}$ is the recorded MR signal, $X_k$ is the design matrix composed of the haemodynamic response, experimental conditions and physiological covariates and $\epsilon_{ik} \sim N(0,\sigma_{ik}^2 V_k)$.  The parameter vector is given by the generalised least squares solution after pre-whitening of the data.  The correlation matrix $V_k$ can be estimated but must be spatially regularised to avoid biased estimates.  The low-level model, therefore, is similar to the active component of model 5 but we have, additionally, estimated the haemodynamic response.

In the multi-level group study approach, the parameter vector, $\beta_{ik}$, for each subject is taken as the response variable in the between-subject model, $[\beta_{i1}^T \cdots \beta_{iK}^T]^T = X_g \beta_g + \epsilon_g$ where $X_g$ is the design matrix composed of subject level covariates such as gender, age, health condition etc.  No frequentist solution to these equations exist if the variance components are unknown so the summary-statistics approach is to solve the low-level model for each subject first and then solve the high-level model based on these estimates (Mumford and Nichols, 2006).

Theoretically, this model could be applied to single-trial variability analysis by estimating the parameter vector for each voxel, $\beta_i$, at each trial and then applying the high-level model to analyse the between-trial variability.  In other words, consider more fixed effects.  This approach may not yield robust estimates given the limited data and high number of parameters unless the high-level model was applied to different sessions of the experiment with the same subject.  This, however, requires the same subject to participate in the experiment multiple times to investigate consistent within-session variability (Liou et al., 2006) which may not be practically, or ethically, possible.

Our model, however, could easily be extended to cope with multiple sessions or subjects.  At present, we analyse the PC scores by looking for trends across voxels within a cluster.  This is not an optimal approach because neighbouring voxels are likely to be correlated but with more runs or subjects, we could invoke a voxel-by-voxel approach to the PC analysis.  It may also, depending on the application, be appropriate to treat the epoch number as a continuous variable and model the PC scores with a more general linear model.  This approach was not implemented here due to the limited number of epochs of each experimental condition and it would not reveal interesting non-linear changes.

Our lower-level model could also be adapted for multiple sessions or subjects.  One variation would be to incorporate a different haemodynamic response function, $h$, for each subject or experimental condition.  We would suggest, however, that including more basis functions would generate a larger number of possible contrasts and make physiological interpretation harder.  The current model captures the inter-trial variability in the covariance model alone.

It remains unclear, however, if the observed single-trial variability is caused by changes in neuronal processes or by natural variation in the reaction time of the volunteer and the strength of the button press.  The amount of noise in the data could be reduced with the inclusion of physiological data as covariates but these data were unavailable for the experimental results presented here.  A further refinement of this model might also include an underlying continuous space/time model, using the real time and location that each slice was recorded and, therefore, eliminating the need for image preprocessing.

The experiment paradigm in our application, with single and multiple button presses, also raised interesting questions regarding the relationship between neuronal activity and haemodynamic response.  It is well documented that this relationship is non-linear, but predicting the change in response to a longer period of activation, for example, is less well understood (Mechelli et al., 2001).  SPM2 adjusts its basis function by convolving the period of neuronal activity with its standard haemodynamic response function, using the Volterra series (Friston et al., 1998).  Independently, the balloon/Windkessel dynamical model of how the haemodynamic response is influenced by the underlying physical changes in the blood vessels was developed (Buxton et al., 1998).  The model suggests that increased blood flow inflates a venous ``balloon", diluting and expelling deoxygenated blood causing an increase in the BOLD signal.  As the flow decreases, the balloon deflates reducing the discharge and increasing the concentration of deoxygenated blood, causing the post-stimulus undershoot.  These models have been shown to be consistent, in that the Volterra kernels which best represented those derived from empirical evidence, also had biologically plausible estimates for the balloon/Windkessel model parameters (Friston et al., 2000).  In this study, we have developed a model that distinguishes between different responses, caused by different neuronal activity, without making assumptions regarding the underlying non-linear relationship.

The methods and results given in this paper obviously pertain to one experiment carried out with one volunteer.  It would be interesting in further work to apply the model and methodology to other volunteers, or repeat it with the same volunteer, to examine if the sources of trial variability found in this study are common to all subjects or experiment repetitions.  If this proves to be the case, it will greatly enhance the ability for neurologists to reach conclusions on voxel activation where a traditional repeated trial experiment paradigm is either impossible to conduct or impractical with the resources available.

To conclude, in this paper we have proposed a statistical model to examine single-trial variability for analysis of ultra-high field fMRI.  This model is an extension of the SPM2 model which is very popular in the literature.  Importantly, through examining the PC scores, we are able to show how the response changes with the task and through time and demonstrate that these changes are statistically significant.  The model also estimates the shape of the mean response and, for the case study considered, is similar to that of prior studies but displays a greater undershoot than previously reported, particularly when the peak is stronger and wider in the cases where the volunteer presses the button multiple times.

{\bf Acknowledgement}

The support of an EPSRC doctoral training account is acknowledged.

{\bf References}
\begin{description}
\item
Aguirre, G.K., Zarahn, E. and D'Esposito, M. (1998). The variability of human BOLD hemodynamic responses. {\it NeuroImage} {\bf 8}, 360-369.
\item
Akaike, H. (1974). A new look at the statistical model identification. {\it IEEE T. Automat. Contr.} {\bf 19}, 716-723.
\item
Auffermann, W.F., Ngan, S.C., Sarkar, S., Yacoub, E. and Hu, X. (2001). Nonadditive two-way ANOVA for event-related fMRI data analysis. {\it NeuroImage} {\bf 14}, 406-416.
\item
Beckmann, C.F., Jenkinson, M. and Smith, S.M. (2003). General multilevel linear modeling for group analysis in fMRI. {\it Neuroimage} {\bf 20}, 1052-1063.
\item
Beckmann, C.F. and Smith, S.M. (2004). Probabilistic independent component analysis for functional magnetic resonance imaging. {\it IEEE T. Med. Imaging} {\bf 23}, 137-152.
\item
Benjamini, Y. and Hochberg, Y. (1995). Controlling the false discovery rate: A practical and powerful approach to multiple testing.
{\it J. Roy. Stat. Soc. B} {\bf 57}, 289-300.
\item
Benjamini, Y. and Hochberg, Y. (2000). On the adaptive control of the false discovery rate in multiple testing with independent statistics. {\it J. Educ. Behav. Stat.} {\bf 25}, 60-83.
\item
Buxton, R.B., Wong, E.C. and Frank, L.R. (1998). Dynamics of blood flow and oxygenation changes during brain activation: The balloon model. {\it Magnet. Reson. Med.} {\bf 39}, 855-864.
\item
Dale, A.M. and Buckner, R.L. (1997). Selective averaging of rapidly presented individual trials using fMRI. {\it Hum. Brain Mapp.} {\bf 5}, 329-340.
\item
Day, N.E. (1969). Estimating the components of a mixture of normal distributions. {\it Biometrika} {\bf 56}, 463-474.
\item
Debener, S., Ullsperger, M., Siegel, M., Fiehler, K., Yves von Cramon, D. and Engel, A.K. (2005). Trial-by-trial coupling of concurrent electroencephalogram and functional magnetic resonance imaging identifies the dynamics of performance monitoring. {\it J. Neurosci.} {\bf 25}, 11730-11737.
\item
Dempster, A.P., Laird, N.M. and Rubin, D.B. (1977). Maximum likelihood from incomplete data via the EM algorithm. {\it J. Roy. Stat. Soc. B} {\bf 39}, 1-38.
\item
Duann, J-R., Jung, T-P., Kuo, W-J., Yeh, T-C., Makeig, S., Hsieh, J-C., Sejnowski, T.J. (2002). Single-trial variability in event-related BOLD signals. {\it NeuroImage} {\bf 15}, 823-835.
\item
Eysenck, H.J. and Frith, C.D. (1977). {\it Reminiscence, motivation and personality}. Plenum, New York.
\item
Friston, K.J., Holmes, A.P., Poline, J-B., Grasby, P.J., Williams, S.C.R., Frackowiak, R.S.J. and Turner, R. (1995a). Analysis of fMRI time-series revisited. {\it NeuroImage} {\bf 2}, 45-53.
\item
Friston, K.J., Holmes, A., Poline, J-B., Price, C.J. and Frith, C.D. (1996). Detecting activations in PET and fMRI: Levels of inference and power. {\it NeuroImage} {\bf 4}, 223-235.
\item
Friston, K.J., Holmes, A.P., Worsley, K.J., Poline, J.-P., Frith, C.D. and Frackowiak, R.S.J. (1995b). Statistical parametric maps in functional imaging: A general linear approach. {\it Hum. Brain Mapp.} {\bf 2}, 189-210.
\item
Friston, K.J., Jezzard, P. and Turner, R. (1994). Analysis of functional MRI time series. {\it Hum. Brain Mapp.} {\bf 1}, 153-171.
\item
Friston, K.J., Josephs, O., Ross, G. and Turner, R. (1998). Nonlinear event-related responses in fMRI. {\it Magnet. Reson. Med.} {\bf 39}, 41-52.
\item
Friston, K.J., Mechelli, A., Turner, R. and Price, C.J. (2000). Nonlinear responses in fMRI: The balloon model, Volterra kernels, and other hemodynamics. {\it NeuroImage} {\bf 12}, 466-477.
\item
Friston, K.J., Penny, W., Phillips, C., Kiebel, S., Hinton, G. and Ashburner, J. (2002). Classical and Bayesian inference in neuroimaging: Theory. {\it NeuroImage} {\bf 16}, 465-483.
\item
Genovese, C.R. (2000). A Bayesian time-course model for functional magnetic resonance imaging data. {\it J. Amer. Stat. Assoc.} {\bf 95}, 691-703.
\item
Glover, G.H. (1999). Deconvolution of impulse response in event-related BOLD fMRI. {\it NeuroImage} {\bf 9}, 416-429.
\item
Goutte, C., Nielsen, F.A. and Hansen, L.K. (2000). Modeling the haemodynamic response function in fMRI using smooth FIR filters. {\it IEEE T. Med. Imaging} {\bf 19}, 1188-1201.
\item
Hansen, L.K., Larsen, J., Nielsen, F.A., Strother, S.C., Rostrup, E., Savoy, R., Lange, N., Sidtis, J., Svarer, C. and Paulson, O.B. (1999). Generalizable patterns in neuroimaging: How many principal components?. {\it NeuroImage} {\bf 9}, 534-544.
\item
Hinrichs, H., Scholz, M., Tempelmann, C., Woldorff, M.G., Dale, A.M. and Heinze, H.J. (2000). Deconvolution of event-related fMRI responses in fast-rate experimental designs: tracking amplitude variations. {\it J. Cog. Neurosci.} {\bf 12}, S76-S89.
\item
Hyvarinen, A., Karhunen, J. and Oja, E. (2001). {\it Independent Components Analysis}. Wiley, New York.
\item
Josephs, O., Turner, R. and Friston, K. (1997). Event-related fMRI. {\it Hum. Brain Mapp.} {\bf 5}, 243-248.
\item
Lindquist, M.A. and Wager, T.D. (2007). Validity and power in hemodynamic response modeling: A comparison study and a new approach. {\it Hum. Brain Mapp.} {\bf 28} 764-784.
\item
Liou, M., Su, H.R., Lee, J.D., Aston, J.A.D., Tsai, A.C. and Cheng, P.E. (2006). A method for generating reproducible evidence in fMRI studies. {\it NeuroImage} {\bf 29} 383-395.
\item
Lu, Y., Jiang, T. and Zang, Y. (2005). Single-trial variable model for event-related data analysis. {\it IEEE T. Med. Imaging} {\bf 24}, 236-245.
\item
Marchini, J.L. and Ripley, B.D. (2000). A new statistical approach to detecting significant activation in function MRI. {\it NeuroImage} {\bf 12}, 366-380.
\item
Mechelli, A., Price, C.J. and Friston, K.J. (2001). Nonlinear coupling between evoked rCBF and BOLD signals: A simulation study of hemodynamic responses. {\it NeuroImage} {\bf 14}, 862-872.
\item
Mumford, J.A. and Nichols, T. (2006). Modeling and Inference of multisubject fMRI data. {\it IEEE Eng. Med. Biol. Magaz.} {\bf 25}, 42-51.
\item
Ward, G., Roberts, M.J. and Phillips, L.H. (2001). Task-switching costs, Stroop-costs, and executive control: A correlational study.
{\it Q. J. Exp. Psychol. A} {\bf 54}, 491-511.
\item
Woolrich, M.W., Behrens, T.E.J., Beckmann, C., Jenkinson, M. and Smith, S.M. (2004a). Multilevel linear modelling for FMRI group analysis using Bayesian inference. {\it Neuroimage} {\bf 21}, 1732-1747.
\item
Woolrich, M.W., Behrens, T.E.J. and Smith, S.M. (2004b). Constrained linear basis sets for HRF modelling using variational Bayes. {\it Neuroimage} {\bf 21}, 1748-1761.
\item
Worsley, K.J. and Friston, K.J. (1995). Analysis of fMRI time-series revisited - again. {\it NeuroImage} {\bf 2}, 173-181.
\item
Worsley, K.J., Liao, C.H., Aston, J., Petre, V., Duncan, G.H., Morales, F. and Evans, A.C. (2002). A general statistical analysis for fMRI data. {\it NeuroImage} {\bf 15}, 1-15.
\item
Zarahn, E. (2002). Using larger dimensional signal subspaces to increase sensitivity in fMRI time series analyses. {\it Hum. Brain Mapp.} {\bf 17}, 13-16.
\end{description}

{\bf Appendix}

We estimate the model parameters in turn, maximising $Q$ conditional on the current values of the other parameters by setting the derivative of $Q$ with respect to the parameter equal to zero and solving for the parameter.  Firstly, we consider the proportion of active voxels, $p$.
\begin{eqnarray*}
\frac{\partial Q}{\partial p} = \sum_{i=1}^V \left\{ \frac{p_i}{p} - \frac{(1-p_i)}{(1-p)} \right\} &=& 0,\\
\Longrightarrow \hspace{3cm} \hat{p} &=& \frac{1}{V} \sum_{i=1}^V p_i.
\end{eqnarray*}
For the scale parameter at each voxel, $i=1,...,V$,
\begin{eqnarray*}
\frac{\partial Q}{\partial \beta_i} = - p_i \left(\beta_i \mu^T \Sigma_1^{-1} \mu - \mu^T \Sigma_1^{-1} (Y_i - X b_i) \right) &=& 0, \\
\Longrightarrow \hspace{1cm} p_i \mu^T \Sigma_1^{-1} \left(\mu \beta_i - (Y_i - X b_i) \right) &=& 0.
\end{eqnarray*}
Hence,
\begin{eqnarray*}
\hat{\beta}_i = (\mu^T \Sigma_1^{-1} \mu)^{-1} \mu^T \Sigma_1^{-1} (Y_i - X b_i).
\end{eqnarray*}
For the parameter vector,
\begin{eqnarray*}
\frac{\partial Q}{\partial b_i} = p_i X^T \Sigma_1^{-1} \left( X b_i - (Y_i - \mu \beta_i) \right) + (1 - p_i) X^T \Sigma_2^{-1} (X b_i - Y_i) = 0.
\end{eqnarray*}
Hence,
\begin{eqnarray*}
\hat{b}_i = (A^T A)^{-1} A^T \left( p_i X^T \Sigma_1^{-1} (Y_i - \mu \beta_i) + (1 - p_i) X^T \Sigma_2^{-1} Y_i \right),
\end{eqnarray*}
where $A = p_i X^T \Sigma_1^{-1} + (1 - p_i) X^T \Sigma_2^{-1}$.  Let $Y_{ij}$ be the response from voxel $i$ at event $j$, $X_j$ the rows of $X$ corresponding to event $j$, and let $e_{jk}$ be the entry at the $j$th row and $k$th column of $\Sigma_E^{-1}$ then differentiating $Q$ with respect to $h$ gives,
\begin{eqnarray*}
\frac{\partial}{\partial h} \left\{ \sum_{i=1}^V \sum_{j=1}^E \sum_{k=1}^E p_i e_{jk} (Y_{ij} - h \beta_i - X_j b_i) \Sigma_T^{-1} (Y_{ik} - h \beta_i - X_k b_i) \right\} &=& 0,\\
\sum_{i=1}^V \sum_{j=1}^E \sum_{k=1}^E p_i e_{jk} \Sigma_T^{-1} \left( \beta_i^2 h -\beta_i (Y_{ij} - X_j b_i) \right) &=& 0.
\end{eqnarray*}
Hence,
\begin{eqnarray*}
\hat{h} = \frac{ \left(\sum_i \sum_j \sum_k p_i e_{jk} \beta_i (Y_{ij} - X_j b_i) \right) }{ \left( \sum_i p_i \beta_i^2 \right) \left( \sum_j \sum_k e_{jk} \right) }.
\end{eqnarray*}
The relative sizes of $\beta_i$ and $\mu = 1_E \otimes h$ are arbitrary so we can artificially rescale them at each iteration such that $\| h \| = 1$ without changing the value of the likelihood.  Let $R_i$ be a $T \times E$ matrix of residuals, where the $j$th column is $Y_{ij} - h \beta_i - X_j b_i$, and let $S_i = R_i \Sigma_E^{-1} R_i^T$.  Then,
\begin{eqnarray*}
\frac{\partial Q}{\partial \Sigma_T^{-1}} = \frac{\partial}{\partial \Sigma_T^{-1}} \left\{ \sum_{i=1}^V p_i E \log | \Sigma_T^{-1} | - p_i \textrm{tr} \left[ \Sigma_T^{-1} S_i \right] \right\} &=& 0.
\end{eqnarray*}
Given, $\partial \log |\Sigma_T^{-1}| / \partial \Sigma_T^{-1} = 2 \Sigma_T - \textrm{Diag} (\Sigma_T)$, and $\partial \textrm{tr} (\Sigma_T^{-1} S) / \partial \Sigma_T^{-1} = 2 S - \textrm{Diag} (S)$, if $\Sigma_T$ and $S$ are symmetric, then,
\begin{eqnarray*}
\sum_{i=1}^V \left\{ p_i E \left[2 \Sigma_T - \textrm{Diag} (\Sigma_T) \right] - p_i \left[2 S_i - \textrm{Diag} (S_i) \right] \right\} &=& 0,\\
\left[2 \Sigma_T - \textrm{Diag} (\Sigma_T) \right] - \frac{1}{E \sum_{i=1}^V p_i} \sum_{i=1}^V p_i \left[2 S_i - \textrm{Diag} (S_i) \right] &=& 0.
\end{eqnarray*}
Let $M = \Sigma_T - \frac{ \sum_i p_i S_i}{E \sum_i p_i}$, then we require $2 M - \textrm{Diag} (M) = 0$.  Hence $M=0$ and,
\begin{eqnarray*}
\hat{\Sigma}_T = \frac{1}{E \sum_{i=1}^V p_i} \sum_{i=1}^V p_i S_i.
\end{eqnarray*}
By analogy, let $T_i = R_i^T \Sigma_T^{-1} R_i$ then,
\begin{eqnarray*}
\hat{\Sigma}_E = \frac{1}{T \sum_{i=1}^V p_i} \sum_{i=1}^V p_i T_i.
\end{eqnarray*}
Finally,
\begin{eqnarray*}
\frac{\partial Q}{\partial \sigma^2} = \sum_{i=1}^V \left\{ \frac{-(1-p_i)ET}{2\sigma^2} + \frac{(1-p_i)}{2\sigma^4} (Y_i - X b_i)^T (Y_i - X b_i) \right\} &=& 0.
\end{eqnarray*}
Hence,
\begin{eqnarray*}
\hat{\sigma}^2 = \frac{1}{ET \sum_{i=1}^V (1-p_i)} \sum_{i=1}^V (1-p_i) (Y_i - X b_i)^T (Y_i - X b_i).
\end{eqnarray*}

\end{document}